\documentstyle[aps,prl,multicol]{revtex}
\begin{document}
\input psfig.sty
%\draft
\title{Precision Determination of the Neutron Spin Structure Function $g^n_1$}

\author{
K. Abe$^{21}$, T. Akagi$^{18, 21}$, B. D. Anderson$^{7}$, 
P. L. Anthony$^{18}$, R. G. Arnold$^{1}$, T. Averett$^{5}$, H. R. Band$^{23}$, 
C. M. Berisso$^{9}$, P. Bogorad$^{15}$, H. Borel$^{6}$, 
P. E. Bosted$^{1}$, V. Breton$^{3}$, M. J. Buenerd$^{18}$, G. D. Cates$^{15}$,
T. E. Chupp$^{10}$, S. Churchwell$^{9}$, K. P. Coulter$^{10}$, 
M. Daoudi$^{18}$, P. Decowski$^{17}$, R. Erickson$^{18}$, 
J. N. Fellbaum$^{1}$, H. Fonvieille$^{3}$, R. Gearhart$^{18}$, 
V. Ghazikhanian$^{8}$, K. A. Griffioen$^{22}$, R. S. Hicks$^{9}$, 
R. Holmes$^{19}$, E. W. Hughes$^{5}$, G. Igo$^{8}$, S. Incerti$^{3}$, 
J. R. Johnson$^{23}$, W. Kahl$^{19}$,
M. Khayat$^{7}$, Yu. G. Kolomensky$^{9}$, S. E. Kuhn$^{13}$, K. Kumar$^{15}$,
M. Kuriki$^{21}$, R. Lombard-Nelsen$^{6}$, D. M. Manley$^{7}$, 
J. Marroncle$^{6}$, T. Maruyama$^{18}$, T. Marvin$^{16}$, W. Meyer$^{4}$,
Z.-E. Meziani$^{20}$, D. Miller$^{12}$, G. Mitchell$^{23}$, 
M. Olson$^{7}$, G. A. Peterson$^{9}$, G. G. Petratos$^{7}$, 
R. Pitthan$^{18}$, R. Prepost$^{23}$, P. Raines$^{14}$, B. Raue$^{13}$,
D. Reyna$^{1}$,
L. S. Rochester$^{18}$, S. E. Rock$^{1}$, M. V. Romalis$^{15}$, 
F. Sabatie$^{6}$, G. Shapiro$^{2}$, J. Shaw$^{9}$, 
T. B. Smith$^{10}$, L. Sorrell$^{1}$,
P. A. Souder$^{19}$, F. Staley$^{6}$, S. St. Lorant$^{18}$, 
L. M. Stuart$^{18}$, F. Suekane$^{21}$, Z. M. Szalata$^{1}$, 
Y. Terrien$^{6}$, A. K. Thompson$^{11}$, T. Toole$^{1}$, X. Wang$^{19}$,
J. W. Watson$^{7}$, R. C. Welsh$^{10}$, F. Wesselmann$^{13}$, T. Wright$^{23}$,
C. C. Young$^{18}$, 
B. Youngman$^{18}$, H. Yuta$^{21}$, W.-M. Zhang$^{7}$, and 
P. Zyla$^{20}$}

\address{
$^{1}$ American University, Washington, DC 20016  \\
$^{2}$ University of California, Berkeley, CA 94720 \\
$^{3}$ LPC IN2P3/CNRS, Univ. Blaise Pascal, F-63170 Aubiere Cedex, France \\
$^{4}$ University of Bonn, Nussallee 12 D-5300 Bonn, Germany \\
$^{5}$ California Institute of Technology, Pasadena, California 91125 \\
$^{6}$ DAPNIA, Saclay, 91191 Gif-sur-Yvette Cedex, France \\
$^{7}$ Kent State University, Kent, OH  44242 \\
$^{8}$ University of California, Los Angeles, CA  90024-1547 \\
$^{9}$ University of Massachusetts, Amherst, MA  01003 \\
$^{10}$ University of Michigan, Ann Arbor, MI  48109 \\
$^{11}$ National Institute of Standards and Technology, Gaithersburg, MD  20899 \\
$^{12}$ Northwestern University, Evanston, IL  60201 \\
$^{13}$ Old Dominion University, Norfolk, VA  23529 \\
$^{14}$ University of Pennsylvania, Philadelphia, PA 19104-6317 \\
$^{15}$ Princeton University, Princeton, NJ  08544 \\
$^{16}$ Southern Oregon State College, Ashland, OR  97520 \\
$^{17}$ Smith College, Northampton, MA 01063 \\
$^{18}$ Stanford Linear Accelerator Center, Stanford, CA  94309 \\
$^{19}$ Syracuse University, Syracuse, NY  13210 \\
$^{20}$ Temple University, Philadelphia, PA  19122 \\
$^{21}$ Tohoku University, Aramaki Aza Aoba, Sendai, Miyagi, Japan \\
$^{22}$ College of William and Mary, Williamsburg, VA  23187 \\
$^{23}$ University of Wisconsin, Madison, WI  53706
}
%\date{}
\maketitle
\vskip 12pt
\centerline{(E154 Collaboration)}
\vskip 12pt
\begin{abstract}
We report on a precision measurement of the neutron spin structure
function $g^n_1$ using deep inelastic scattering of polarized electrons by 
polarized $^3$He.
For the kinematic range 0.014$<x<$0.7 and 1 (GeV/c)$^2<$ 
${\it Q}^2<$ 17 (GeV/c)$^2$,
we obtain
$\int^{0.7}_{0.014}\ g^n_1(x)dx$=$-$0.036 $\pm$ 0.004 (stat) $\pm$
0.005 (syst) at an average $Q^2$=5 (GeV/c)$^2$.  
We find relatively large negative values for $g^n_1$
at low $x$.  The results call into question the usual Regge theory method 
for extrapolating
to $x$=0 to find the full neutron integral $\int^1_0 g^n_1(x)dx$,
needed for testing quark-parton model and QCD sum rules. 
\end{abstract}
\begin{multicols}{2}[]
\narrowtext

Deep inelastic scattering (DIS) of polarized leptons by polarized nucleons
has been the cornerstone
for studying the internal spin structure of the proton and neutron.  
Although the first experiments\cite{E80:76,E130:83} found large 
asymmetries in the spin-dependent scattering
of electrons by protons, consistent 
with the early quark-parton
model (QPM) predictions\cite{Kaur:77}, subsequent 
experiments\cite{EMC:88,SMC:94,E143:95} 
performed at higher energies
found that the proton asymmetries at low values of
Bjorken $x$ disagreed with the early QPM predictions.  In fact,  
higher energy proton measurements were inconsistent with
one of the QPM sum rules
derived by Ellis and Jaffe\cite{EJ:74} based upon an unpolarized strange sea.
First measurements of spin-dependent scattering of polarized leptons 
off polarized
neutrons found small negative asymmetries, and, along with
the proton results, provided the first tests of the fundamental
Bjorken Sum Rule\cite{BJ:66}.  
However, the neutron results suffered either from large
statistical uncertainties at low $x$\cite{SMC:95,SMCn2:97}, or 
from a limited beam
energy\cite{E142:93,E143n:95}.
This Letter reports on a precision measurement of the neutron spin structure
function 
$g_1^n$ performed at the Stanford Linear Accelerator 
Center (SLAC) using 48.3 GeV polarized
electrons scattered from polarized $^3$He to achieve $x$
values as low as 0.014.  The present experiment (E154), which collected
$10^8$ events in October and November of 1995, builds on
the experience from the previous
SLAC $^3$He experiment (E142)\cite{E142:93} performed at a lower beam energy.  
The E154 results provide a
new insight into the low $x$ behavior of $g_1^n$.

The asymmetries $A_{\parallel}$($A_{\perp}$) measured in DIS
of longitudinally polarized electrons by longitudinally
(transversely) polarized nucleons can be used to find the
nucleon spin structure function $g_1$\cite{PRD:96}, namely
\end{multicols}
\widetext

$$g_1(x,Q^2) = F_2(x,Q^2){1 + \gamma^2 \over 2xD'(1 + R(x,Q^2))}
[A_{\parallel} + {\rm tan}(\theta/2) A_{\perp}],$$

\begin{multicols}{2}[]
\narrowtext

\noindent
where $Q^2$ is the squared four-momentum transfer of the virtual photon; 
$x$ is the fraction of nucleon
momentum carried by the struck quark;
$\gamma$ and $D'$  
are
factors depending on the scattered electron's initial and final
energies and the electron scattering angle $\theta$; 
$F_2(x,Q^2)$ is the unpolarized nucleon spin
structure function and 
$R$($x$, $Q^2$) = $\sigma_L / \sigma_T$ is the longitudinal to transverse
virtual photoabsorption cross section ratio.
The asymmetries $A_{\parallel}$($A_{\perp}$) may also be used to find the 
virtual photon-nucleon asymmetries $A_1(x,Q^2)$.

Polarized electrons were obtained using
a strained GaAs cathode illuminated by circularly polarized
light from a flashlamp-pumped Ti:sapphire laser\cite{Source:92}.  
The electron spin direction was reversed randomly
on a pulse-to-pulse basis by reversing the helicity of the laser light.
The electrons were subsequently accelerated to 48.3 GeV and directed
to the experimental hall.
The charge per pulse ranged from 3 to
9 x 10$^{10}$ electrons, yielding an average current ranging from
0.5 to 2 $\mu$A for a pulse repetition rate of 120 Hz and a pulse width
of 250 ns.
The beam polarization was measured to be
0.82 $\pm$ 0.02 over the duration of the experiment using                  
a single arm M\o ller polarimeter\cite{Mol:97}
located upstream of the target. 

The polarized $^3$He target
consisted of double-chamber glass cells\cite{chupp2:92}
filled with $\sim$9.5 atm 
of $^3$He (as measured at 20$^{\circ}$C).  The 30 cm long cells 
were constructed of Corning 1720 glass.  The lower chamber
had $\sim$50 $\mu$m inverted end windows through 
which the electron beam passed.
Approximately 50 torr of nitrogen
gas was also present in the cells to aid in optical pumping. 
The $^3$He nuclei were
polarized in the upper chamber by spin-exchange collisions with 
optically-pumped polarized rubidium atoms\cite{Bou:60,Chupp:87}.  
Three 20 W 
diode lasers and four Argon-ion pumped
Ti:sapphire lasers continuously polarized the
rubidium atoms in the upper chamber of
the target cell.  The target spin direction was reversed 
approximately once a week throughout
the experiment.
NMR techniques\cite{NMR:61} calibrated by proton NMR and by frequency 
shift techniques\cite{Princ:93}, were used to measure the
polarization of the $^3$He nuclei.  The polarization ranged as high
as 0.48 and was on average
0.38 $\pm$ 0.02 over the duration of the experiment.
The systematic uncertainty in the target polarization was dominated by 
the water calibration for the NMR
technique and by uncertainties in the polarization gradients
and $^3$He density for the frequency
shift technique.

Two new single-arm spectrometers, at
central scattering angles of 2.75$^{\circ}$ and 5.5$^{\circ}$, 
were used to analyze scattered electrons\cite{YURY:97}.
Each spectrometer
utilized a pair of threshold \v Cerenkov counters operating with nitrogen at
a pressure of 0.10 (0.14) atm in the 2.75$^{\circ}$ (5.5$^{\circ}$) arm,
corresponding to a pion energy threshold of
approximately 19 (16) GeV.  Ten (eight) planes of hodoscopes were used for
tracking in the 2.75$^{\circ}$ (5.5$^{\circ}$) spectrometer.  Tracking
resolution resulted in a momentum determination ranging from $\pm$ 2\% at 
low momentum
to $\pm$ 4\% at high momentum.
The momentum resolution was useful for reducing
the contamination from hadronic backgrounds to 
the electron sample.  At the rear
of each spectrometer a 200 block lead glass calorimeter was
arranged in a fly's eye configuration\cite{French:95} which gave an
energy resolution
of 3\% + (8/$\sqrt{E(\rm{GeV})}$ )\%.  Only events 
with scattered electron energies greater than 10 GeV were used
in the analysis, corresponding to $Q^2 >\ $ 1 (GeV/c)$^2$ for the 
2.75$^\circ$ spectrometer.

For each beam pulse, the experiment collected 
information from the hodoscope and calorimeter multihit
TDCs and the calorimeter ADCs.  
The four \v Cerenkov counters were each read out by a
Flash ADC that recorded the pulse shape in 1 ns time 
slices covering the
full beam pulse.  Events were analyzed as electron candidates if
they passed a low-threshold in both \v Cerenkov
counters in
coincidence with an energy cluster in the lead glass.  
Events were tracked using the
lead glass centroid cluster position and hits in the hodoscope 
planes.  The tracks, combined with information on the spectrometer optics, 
were used to determine the particle's momentum.
Tracking efficiency was measured to be on
the order of 90\%.  
Events were also classified by the energy deposition
in the calorimeter. 
When the ratio of the energy deposited in the calorimeter to the momentum
determined from tracking for an event was less than 80\%, the event was
rejected as a pion candidate.
Typically 0.5 (0.2) electrons and 5 (2) pions were recorded per pulse
in the 2.75$^{\circ}$ (5.5$^{\circ}$) spectrometer.  
Selected events were binned in $x$ and tagged per pulse with 
the relative beam and target spin directions. 
\begin{figure}
\vskip 2.75in
\centerline{\psfig{figure=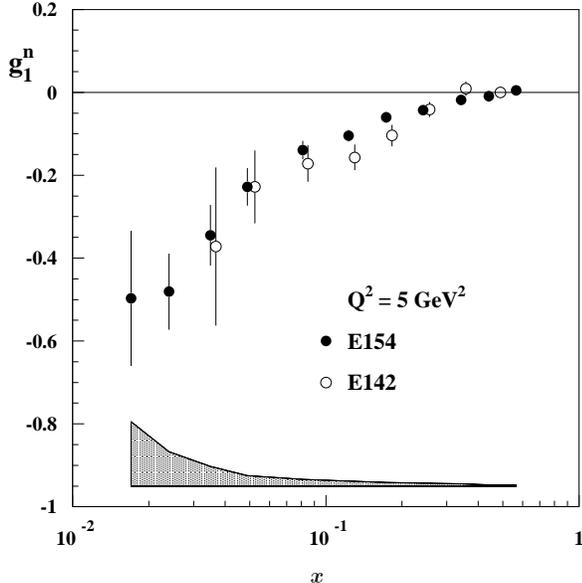,width=3.25in}}
\caption{Results for $g_1^n$ versus $x$ from SLAC Experiment E154
compared to Experiment E142
evaluated at $Q^2$ = 5 (GeV/c)$^2$.  Shaded region corresponds to 1 $\sigma$
systematic uncertainties.}
\end{figure}

Contamination of hadronic background in the electron sample was measured to
be 3 $\pm$ 2 \% for the lowest $x$ values and decreased at higher values.
Furthermore, since the hadron asymmetries were found to be approximately 1/3
the size of the electron asymmetries,  the total effect of hadron
contamination was very small.  On the other hand, a relatively large
contamination of the DIS electron sample originates 
from electrons produced
from charge-symmetric decays of hadrons.
The rates from this background were
determined from running with the spectrometer polarity reversed
to measure positrons.
The rates for the non-DIS
electron event background were on the order of 15\% at the lowest 
scattered electron energies
and fell rapidly with increasing energy.
The measured asymmetries from these runs were found to be consistent
with zero. 

The fraction of DIS events that come from polarized $^3$He as compared to
the full target cell is called the dilution factor.  It was determined
from known unpolarized nucleon structure
functions, measured glass cell window thicknesses and
the density of gas in the target cells (material method).
The dilution factor was also determined by comparing rates from the polarized
target to rates from a dummy cell with different gas pressures (rate method).
This method has the advantage of taking into account possible beam halo
effects.  Results were obtained using the material method, and the rate
procedure was used to assign systematic uncertainties.
On average, the dilution
factor was found to be 0.55 $\pm$ 0.03.

\begin{figure}
\centerline{\psfig{figure=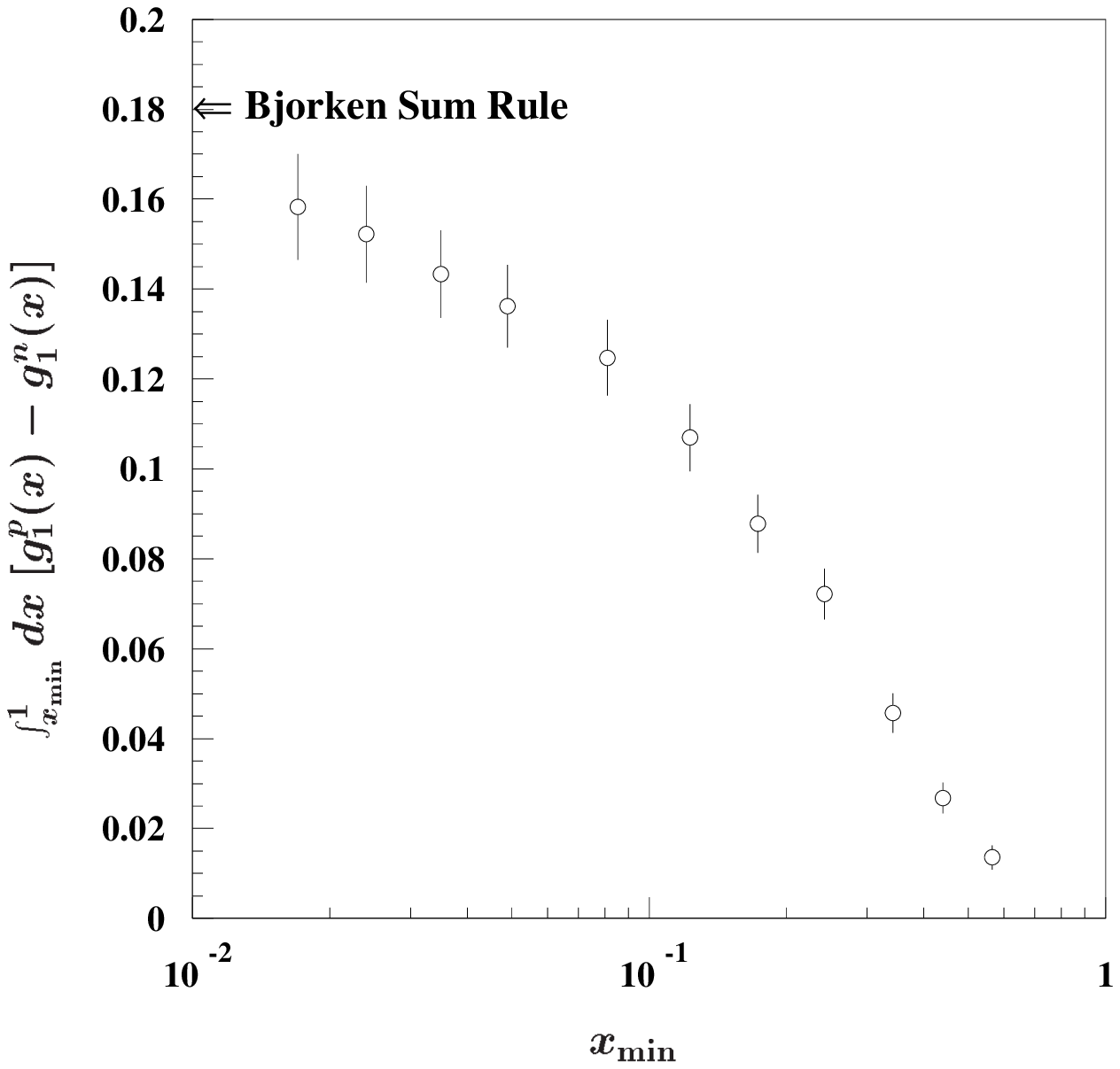,width=4.5in}}
\caption{Difference between the 
measured proton [5,6] and neutron [This experiment] integrals 
calculated from
a minimum $x$ value, $x_{\rm min}$ up to $x$ of 1.  The value is compared to
the theoretical prediction from the Bjorken sum rule which makes a prediction
over the full $x$ range.  For the prediction, the Bjorken sum rule is evaluated
up to third order in $\alpha_s$\protect\cite{Lar:91,PDG:96} and 
at $Q^2$ = 5 (GeV/c)$^2$.
Error bars on the data are dominated by systematic uncertainties
and are highly correlated point-to-point.  }
\end{figure}

After corrections for hadronic and pair-symmetric backgrounds, dilutions and
polarizations, the asymmetries $A_{\parallel}$ and
$A_{\perp}$ were formed.  The
asymmetries were corrected for radiative processes to find the 
single-photon exchange Born results\cite{RC:83,RC:94,TSAI:69,TSAI2:71}. 
Uncertainties in the radiative corrections were estimated
by varying the input models over a range consistent with the measured data. 

Corrections due to the nuclear wave function of the polarized $^3$He nucleus
were applied\cite{Ciofi:93,Wol:84,Friar:90,Sauer:93} 
using the recent proton data\cite{SMC:94,E143:95} to evaluate the 
proton contributions; however these contributions had only a small
impact on the results.  
No other corrections were made for the fact that the polarized neutron
is embedded in the $^3$He nucleus.

Results for $A_1^n$ and 
$g_1^n$ are presented in Table 1, and $g_1^n$ is plotted
in Fig. 1 along with the results of the 
SLAC E142 experiment\cite{E142:93}.  The results from both experiments
are evolved to $Q^2$ = 5 (GeV/c)$^2$
under the assumption that $g_1/F_1$ is independent of $Q^2$.   
Within experimental uncertainties, this assumption is supported by a 
comparison of our data to all existing measurements
\cite{SMC:95,SMCn2:97,E142:93,E143n:95,QS:95,HERM:96}.
Good agreement with the
E142 results is seen in the overlapping $x$ range.  
Over
the range of this experiment, 
we find a neutron spin structure function integral of
$\int_{0.014}^{0.7} g_1^n(x)dx$ = $-$0.036 $\pm$ 0.004 (stat.) 
$\pm$ 0.005 (syst.). 

A notable feature of Fig. 1 is the strong $x$-dependence observed at
low $x$, a result that is incompatible with the simplest Regge theory
interpretation\cite{Hei:73,EK:88} 
that $g_1^n$ is constant with $x$ in this region.  The strong
$x$-dependence also implies that the unmeasured small-$x$ region can make
a major contribution to the integral $\int_0^1 g_1^n(x)dx$ and recourse
must be made to models in order to evaluate the full integral.  The 
result is that the value extracted for the integral is subject to
considerable model uncertainty.  For example, a Regge theory extrapolation
with functional form $g_1^n \sim x^{-\alpha}$, $-0.5 < \alpha \le 0$,
yields $\int_0^1 g_1^n(x)dx$ = $-$0.041 $\pm$ 0.004 $\pm$ 0.006, even though
this description is successful in fitting only the three lowest $x$ points
at $x < 0.04$.
In contrast, a fit to the $x < 0.1$ data with an
unconstrained power-law yields 
$\int_0^1
g_1^n(x)dx$ = $-$0.2. No uncertainty can be given for this later analysis,
since the fitted value of $\alpha$ is 0.9 $\pm$ 0.2, and the integral
diverges for $\alpha$ = 1.
Figure 3 summarizes the 
results of the fits described above to the low $x$ data region.
If we fit an unconstrained power law to the measured $g_1^n$ values 
without evolving to $Q^2$ = 5 GeV$^2$, we find $\alpha$ is 0.7 $\pm$ 0.2. 
In short, the new data do not adequately constrain
the low-$x$ region such that the integral of $g_1^n$ can be reliably 
extracted. 

\begin{figure}
\centerline{\psfig{figure=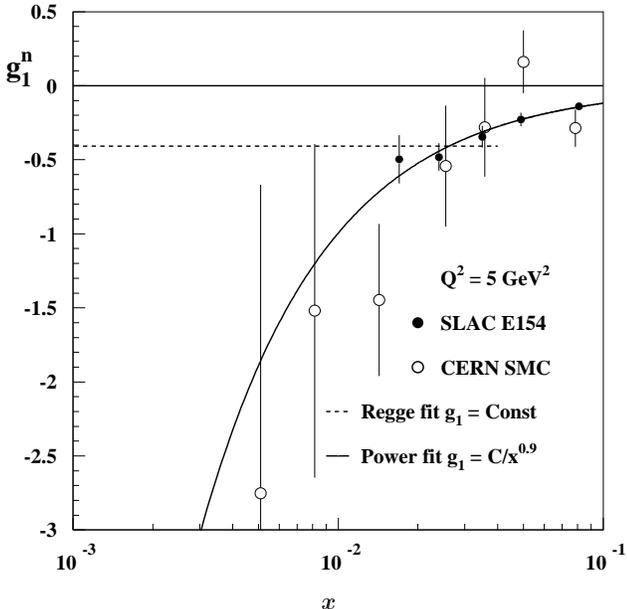,width=3.5in}}
\caption{Results for $g_1^n$ versus $x$ for the low $x$ region 
from SLAC experiment E154 compared
to the CERN SMC experiment.  The data is evolved to $Q^2$ = 5 (GeV$^2$/c$^2$).
Fits that impact the low $x$ extrapolation (discussed in the text) are
presented.}
\end{figure}

We have also used the present precision neutron results down to 
$x_{\rm min}=0.014$
along with the
proton results from SLAC E143 experiment ($0.03 < x$)\cite{E143:95} and the
CERN SMC experiment ($0.014 < x < 0.03$)\cite{SMC:94} to compare 
to the Bjorken sum rule prediction.
  The difference between proton and neutron spin structure
functions integrated over $x$ from $x_{min}$ to $x$=1 is shown in
Fig. 2.  One sees that the difference in the integral of $g_1$ 
for the proton and neutron falls only 1.9 standard deviations
below the Bjorken sum rule prediction when the data is integrated down
to $x$ of 0.014.  Presumably the rest of the integral comes from the
remaining unmeasured low $x$ region. 

In conclusion, we have found relatively large negative values of $g_1^n$ at
low $x$.  One possible explanation for this behavior can be associated with 
sea and gluon
spin contributions\cite{Forte:96,Gluck:96,Gehrmann:96}.
A breakdown in the simple Regge theory description
at low $x$ is also a possible consequence.
Further precision data using proton and deuteron targets
over the same kinematic range are expected to be of great use in unraveling
the behavior of the nucleon spin structure functions at 
moderately low $x$ (down to $x \approx$ 0.01).  High precision low $x$
measurements of the nucleon spin structure functions are 
still needed to understand how $g_1^n$ converges at low $x$ and to extract
the neutron integral $\int_0^1 g_1^n(x)dx$.

We thank the personnel of the SLAC accelerator department for their
efforts which resulted in the successful operation of the E154 Experiment.
This work was supported by the Department of Energy; by the National
Science Foundation;
by the Kent State University Research Council (GGP); 
by the Jeffress Memorial Trust (KAG);    
by the Centre National de la Recherche Scientifique and the
Commissariat a l'Energie Atomique (French groups); and by the Japanese
Ministry of Education, Science and Culture (Tohoku).

\end{multicols}
\widetext

\begin{table}
\begin{center}
\caption{Results on $A_1^n$ and $g_1^n$ 
at the measured $Q^2$, along with  
$g_1^n$  evaluated at 
$Q^2=5~({\rm GeV}/c)^2$ assuming that  
$g_1^n/F_1^n$ is independent of $Q^2$.}
\begin{tabular}{cccccc} 
$x$ range & $\langle~x~\rangle$ & 
$\langle~Q^2~\rangle$ & 
$g_1^n\pm\mathrm{stat.} 
\pm\mathrm{syst.}$& 
$A_1^n\pm\mathrm{stat.} 
\pm\mathrm{syst.}$& 
$g_1^n\pm\mathrm{stat.} 
\pm\mathrm{syst.}$\\ 
 & & $(\mathrm{GeV/c})^2$ & & &   
$ (Q^2 = 5~(\mathrm{GeV}/c)^2$)\\ \hline 
$0.014-0.02$&$0.017$&$\phantom{0}1.2$&                                
$-0.351\pm0.115\pm0.110$&
$-0.058\pm0.019\pm0.018$&
$-0.497\pm0.163\pm0.155$
\\
$0.02\phantom{0}-0.03$&$0.024$&$\phantom{0}1.6$&                      
$-0.374\pm0.071\pm0.065$&
$-0.080\pm0.015\pm0.014$&
$-0.481\pm0.092\pm0.083$
\\
$0.03\phantom{0}-0.04$&$0.035$&$\phantom{0}2.0$&                      
$-0.290\pm0.061\pm0.039$&
$-0.078\pm0.018\pm0.011$&
$-0.345\pm0.073\pm0.047$
\\
$0.04\phantom{0}-0.06$&$0.049$&$\phantom{0}2.6$&                      
$-0.204\pm0.040\pm0.022$&
$-0.086\pm0.016\pm0.010$&
$-0.228\pm0.045\pm0.025$
\\
$0.06\phantom{0}-0.10$&$0.081$&$\phantom{0}4.4$&                      
$-0.137\pm0.021\pm0.016$&
$-0.092\pm0.013\pm0.011$&
$-0.139\pm0.022\pm0.016$
\\
$0.10\phantom{0}-0.15$&$0.123$&$\phantom{0}6.6$&                      
$-0.108\pm0.015\pm0.012$&
$-0.106\pm0.014\pm0.012$&
$-0.105\pm0.014\pm0.012$
\\
$0.15\phantom{0}-0.20$&$0.173$&$\phantom{0}8.2$&                      
$-0.061\pm0.014\pm0.009$&
$-0.092\pm0.021\pm0.012$&
$-0.060\pm0.014\pm0.009$
\\
$0.20\phantom{0}-0.30$&$0.242$&$\phantom{0}9.8$&                      
$-0.042\pm0.011\pm0.007$&
$-0.112\pm0.028\pm0.020$&
$-0.043\pm0.011\pm0.007$
\\
$0.30\phantom{0}-0.40$&$0.342$&$11.7$&                                
$-0.017\pm0.011\pm0.005$&
$-0.068\pm0.065\pm0.025$&
$-0.018\pm0.013\pm0.005$
\\
$0.40\phantom{0}-0.50$&$0.441$&$13.3$&                                
$-0.007\pm0.011\pm0.002$&
$-0.003\pm0.142\pm0.022$&
$-0.009\pm0.014\pm0.003$
\\
$0.50\phantom{0}-0.70$&$0.564$&$15.0$&                                
$\phantom{-}0.003\pm0.008\pm0.001$&
$\phantom{-}0.100\pm0.294\pm0.039$&
$\phantom{-}0.005\pm0.012\pm0.002$
\\
\end{tabular} 
\end{center} 
\end{table} 

\end{document}